\documentclass[conference]{IEEEtran}
\IEEEoverridecommandlockouts
\usepackage{cite}
\usepackage{amsmath,amssymb,amsfonts}
\usepackage{algorithm}
\usepackage{algpseudocode}
\usepackage{graphicx}
\usepackage{textcomp}
\usepackage[dvipsnames]{xcolor}
\usepackage{siunitx}
\usepackage{subcaption}
\usepackage{pifont}
\usepackage{makecell}
\newcommand{\todo}[1]{\textcolor{red}{\bf #1}}
\newcommand{\aj}[0]{\textcolor{blue}}
\usepackage{url}

\usepackage{caption}
\newcommand{\red}[1]{{\color{red}#1}}

\newcommand{\blue}[1]{{\color{blue}#1}}

\begin{document}

\title{Machine Learning for CUDA+MPI Design Rules}

\author{\IEEEauthorblockN{Carl Pearson}
\IEEEauthorblockA{
\textit{Sandia National Laboratories}\\
Albuquerque, NM, USA \\
cwpears@sandia.gov}
\and
\IEEEauthorblockN{Aurya Javeed}
\IEEEauthorblockA{
\textit{Sandia National Laboratories}\\
Albuquerque, NM, USA \\
asjavee@sandia.gov}
\and
\IEEEauthorblockN{Karen Devine}
\IEEEauthorblockA{
\textit{Sandia National Laboratories}\\
Albuquerque, NM, USA \\
kddevin@sandia.gov}
}

\maketitle

\begin{abstract}

We present a new strategy for automatically exploring the design space of key CUDA+MPI programs and providing design rules that discriminate slow from fast implementations.
In such programs, the order of operations (e.g., GPU kernels, MPI communication) and assignment of operations to resources (e.g., GPU streams) makes the space of possible designs enormous.
Systems experts have the task of redesigning and reoptimizing these programs to effectively utilize each new platform.
This work provides a prototype tool to reduce that burden.

In our approach, a directed acyclic graph of CUDA and MPI operations defines the design space for the program.
Monte-Carlo tree search discovers regions of the design space that have large impact on the program's performance.
A sequence-to-vector transformation defines  features for each explored implementation, and each implementation is assigned a class label according to its relative performance.
A decision tree is trained on the features and labels to produce design rules for each class; these rules can be used by systems experts to guide their implementations.
We demonstrate our strategy using a key kernel from scientific computing --- sparse-matrix vector multiplication --- on a platform with multiple MPI ranks and GPU streams.  

\end{abstract}

\begin{IEEEkeywords}
Computer performance, Parallel programming, Parallel machines, Decision support systems, Machine learning, Monte Carlo simulation
\end{IEEEkeywords}

\section{Introduction}

This work presents a new strategy for automatically exploring the design space of key CUDA+MPI operations and providing design rules that discriminate slow from fast implementations.
High performance MPI+CUDA programs are assembled from more primitive operations, particularly GPU kernels, inter-node communication, and GPU data transfers.
In order to overlap communication with computation, these operations are typically asynchronous and parallel. 
The order of these operations and how they are assigned to system resources greatly affect  performance, but optimal orders and assignments can be difficult to identify.

\begin{figure}[ht]
\centering
\includegraphics[width=\linewidth]{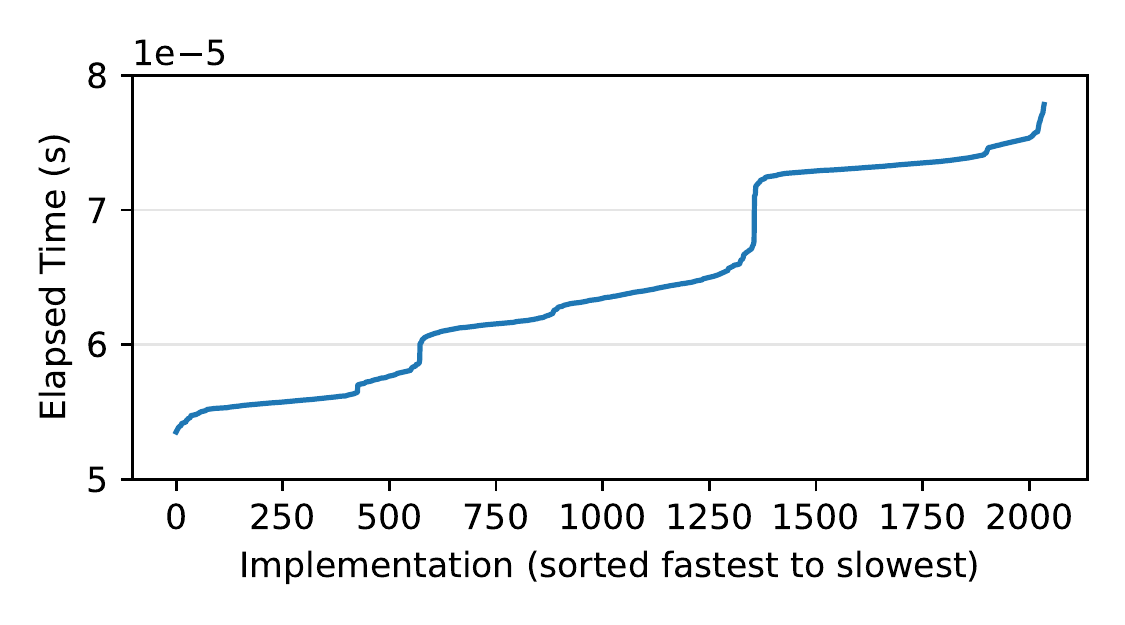}
\caption{
$1.47\times$ speedup between the fastest and slowest of \num{2036} distributed sparse-matrix vector multiplication implementations.
All implementations use the same GPU kernels and MPI functions; only the order of operations and stream assignments are modified.
}
\label{fig:motivation}
\end{figure}

For example, Figure~\ref{fig:motivation} shows the performance of \num{2036} different distributed MPI+CUDA sparse-matrix vector multiplication (SpMV, Section~\ref{sec:design-space-exploration}) implementations sorted from fastest to slowest
The fastest implementation is $1.47\times$ faster than the slowest.
The kernel implementations and MPI communication methods are identical -- only the order of operations and the assignment of GPU operations to streams are changed.

The freedom to reorder independent operations and assign those operations to various system resources yields a combinatorial explosion of possible implementations, even for relatively simple distributed CUDA+MPI programs.
Experts typically design these programs according to various heuristics (e.g. longest possible communication window, maximum GPU utilization, etc.) and then engage in a process of iterative measurement and refinement to achieve their objectives.

Different computer systems have different performance characteristics, forcing implementers to repeat this process for each target system.
The example in Figure~\ref{fig:motivation} is just a single matrix input on a single execution platform.
As new systems with new performance characteristics come online, program authors must revisit the design of their key operations to ensure continued performance.

We simplify this design process with a new system for automatically exploring the design space of  CUDA+MPI programs and providing design rules that discriminate slow from fast implementations.  Our contributions are
\begin{itemize}
\item A Monte-Carlo tree search strategy for exploring the CUDA+MPI design space, and
\item A decision-tree-based method for generating rules to discriminate fast from slow programs.
\end{itemize}

\begin{figure*}[ht]
\centering
\includegraphics[width=\linewidth]{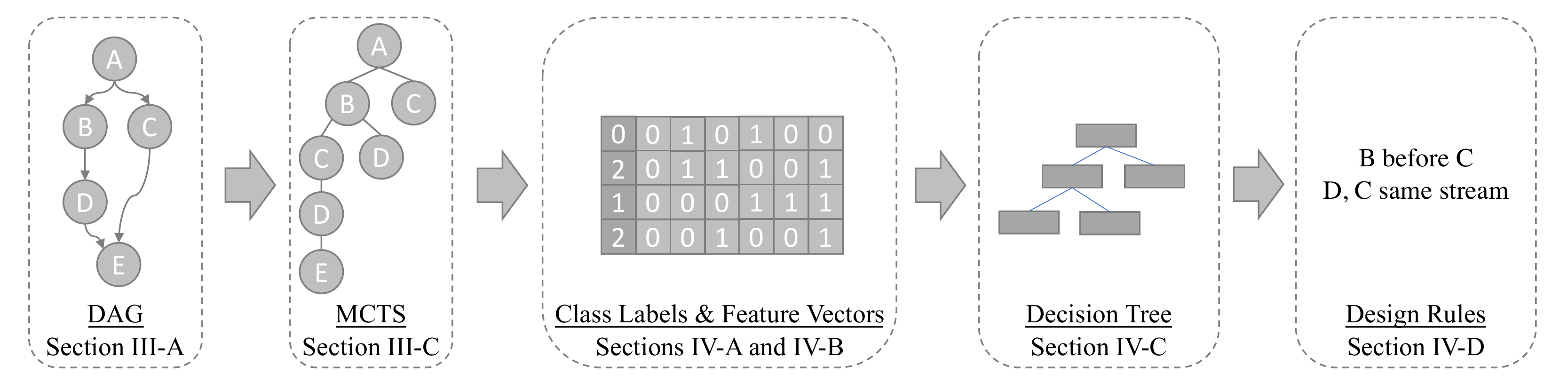}
\caption{
Overview of the proposed system.
A directed acyclic graph of operations defines the design space (Section~\ref{sec:dag}).
Monte-Carlo tree search identifies interesting regions of the design space (Section~\ref{sec:mcts}).
Class labels are generated through convolution and peak detection (Section~\ref{sec:class-labels}).
Feature vectors are created through a custom sequence-to-vector transformation (Section~\ref{sec:feature-vectors}).
A decision tree trained on the features (Section~\ref{sec:tree-training}) and labels produces design rules (Section~\ref{sec:rule-generation}).
}
\label{fig:overview}
\end{figure*}

The goal is to provide program implementers with specific design rules for meeting performance objectives.
Simultaneously, ``black-box'' tuning results are avoided, granting implementers complete provenance over the final program design.

Figure~\ref{fig:overview} outlines the system.
A directed acyclic graph of CUDA and MPI operations defines the design space for the program (Section~\ref{sec:dag}).
Monte-Carlo tree search discovers regions of the design space that have large impact on the program performance (Section~\ref{sec:mcts}).
A sequence-to-vector transformation is applied to define features for each explored implementation, and each implementation is assigned a class label according to its relative performance (Sections~\ref{sec:class-labels} and \ref{sec:feature-vectors}).
A decision tree is trained on the features and labels to produce design rules for each class (Sections~\ref{sec:tree-training} and \ref{sec:rule-generation}).

We demonstrate our strategy with a CUDA+MPI sparse-matrix vector multiplication on the Perlmutter HPE Cray EX platform at NERSC (Sections~\ref{sec:design-rule-generation} and~\ref{sec:mcts-results}).
Long term, the system will be applied to high-performance libraries like Tpetra~\cite{trilinos-website, tpetra-website}, and the design-rule generation will be expanded to accommodate multiple platforms and multiple inputs.
An open-source implementation is available at \url{github.com/sandialabs/tenzing}.

\section{Background}

Both MPI and GPUs typically are used in an asynchronous manner, in which the CPU initiates communication or computation operations and proceeds with its own thread of execution while the MPI implementation and GPUs work in parallel.
Asynchronous operations allow overlapped communication and computation, since (in principle) a GPU kernel can be executing while an MPI communication is in-flight.

CUDA provides streams and events to control and synchronize  asynchronous operations.
A GPU stream is a queue of such events in which each operation must complete before the next begins.
A GPU has multiple streams, allowing multiple sequences of operations to proceed independently (and in parallel when execution resources permit).
A CUDA event represents a particular point in a stream's execution.
It allows the CPU or other streams to synchronize with a particular state of a given stream, either waiting for that state to be realized in the future, or proceeding immediately if that state already existed in the past.

MPI's facility for synchronization is the MPI\_Request.
An MPI\_Request object is associated with a particular MPI operation, e.g. MPI\_Isend.
The CPU can interact with that object to query or control the ongoing communication operation.

\subsection{Monte-Carlo Tree Search For Performance Optimization}

Monte-Carlo tree search~\cite{abramson2014expected, coulom2006efficient} (MCTS) is a heuristic search method for decision problems that describes the search space in a tree.
Each tree node represents a state of the search, and the node's children describe possible next states from the current state.
The tree is grown by balancing exploration and exploitation values for each node, which are used to guide the search towards the optimal result.
MCTS (and the similar beam search) have been examined previously in the context of performance optimization.
As in our work, MCTS has been used for task-scheduling, though it does not explicitly seek the fastest implementation.
Instead, MCTS is used to discover regions of the design space that have a large effect on performance.

Adams et al.~\cite{adams2019learning} use beam search to generate image-processing and deep-learning programs from high-level Halide descriptions.
Wang et al.~\cite{wang2019alphax} use MCTS to design neural network architectures.
Exploitation is based on network accuracy.
The search is slowed by long network training times before accuracy can be evaluated.
A neural network is used to estimate the performance of the proposed networks during MCTS.
Zheng et al.~\cite{zheng2020ansor} propose Ansor for automatically generating tensor programs.
Among other things, Ansor attempts to generate fast task graphs for deep neural networks.
Ansor engages in random sampling of the task graph space, and then uses a gradient-descent approach for scheduling those tasks.
Our work uses MCTS to address sampling and scheduling jointly.
Halaji et al.~\cite{hajali2020protuner} use MCTS to tune deep learning and image processing programs.

MCTS has also been used to determine optimal parameters for various tasks.
Kruse, Finkel, and Wu~\cite{kruse2020autotuning} and Koo et al.~\cite{koo2021customized}  use MCTS for choosing loop optimizations in LLVM's polyhedral loop optimizer.
Anderson et al.~\cite{anderson2020learning} uses beam search to optimize loop nests for Halide programs.
Our work does not use MCTS to optimize individual tasks; rather, in this work, we focus on task scheduling.

\subsection{Decision Trees in Program Optimization}
Decision trees represent decision processes as a binary tree of conditional statements.
Each node in the tree tests a particular feature and each child represents an outcome.
Decision trees have been used in various aspects of program optimization.
As opposed to prior works, we do not use the decision tree to make optimization decisions, but instead to determine distinguishing properties of various implementations.

Monsifrot and Amarasingh~\cite{monsifrot2002machine} use a decision tree to choose whether to unroll program loops based on hand-crafted features, followed by Leather, Bonilla, and O'Boyle~\cite{leather2014automatic} to choose loop unroll factors.
Yu and Rauchwerger~\cite{yu2014adaptive} use decision trees to predict appropriate loop parallelization strategies.
Grewe, Wang, and Boyle \cite{grewe2013portable} and Wang, Grewe, and Boyle~\cite{wang2014automatic} use decision trees to decide how effective GPU acceleration will be.
Ding et al.~\cite{ding2015autotuning} use decision trees to choose the optimal algorithm implementation for different inputs.
Lokuciejewski~\cite{lokuciejewski2009automatic} uses random forests to predict the benefits of function inlining.
Benedict et al.~\cite{benedict2015energy} and Rejitha, Benedict, Alex, and Infanto ~\cite{rejitha2017energy} use random forests to predict energy cost of OpenMP and CUDA programs.

\subsection{MCTS for Scheduling Outside of Computing}
MCTS has been applied to non-computer-science task-scheduling problems as well.
Lubosch, Kunath, and Winkler~\cite{lubosch2018industrial} apply MCTS to a benchmark semiconductor fabrication scheduling problem.
Runarsson, Schoenauer, and Sebag~\cite{runarsson2012pilot} apply MCTS to a job-shop scheduling problem, and find that MCTS produces as good or better results than other algorithms.
Wu, Wu, and Liang~\cite{wu2013multi} similarly consider a multi-objective version of the problem, and determine that MCTS finds state-of-the-art solutions in much less time.
Neto, Constantino, Martins, and Pedroso~\cite{NETO2020multi} apply MCTS to a forestry harvesting problem, where stands of trees are assigned a lifespan before harvesting.
They find that MCTS is able to find solutions for large problems in reasonable time.

\section{Design Space Exploration}
\label{sec:design-space-exploration}

The combination of stream assignments and operation orderings can yield a tremendous number of valid implementations; too many to enumerate, much less explore.
Even for the simple SpMV example described below there are \num{2036} implementations.
Some implementation decisions have a large impact on performance, and some do not.
Figure~\ref{fig:motivation} shows this effect --- certain decisions  move an implementation from one performance regime to another, but many decisions have little effect.
This work uses Monte-Carlo tree search to jointly explore the space of implementations, biasing the search to focus on regions of high performance impact, without needing the space to be enumerated a priori.

\begin{figure*}[ht]
     \centering
     \begin{subfigure}[t]{0.22\textwidth}
         \centering
         \includegraphics[width=\textwidth]{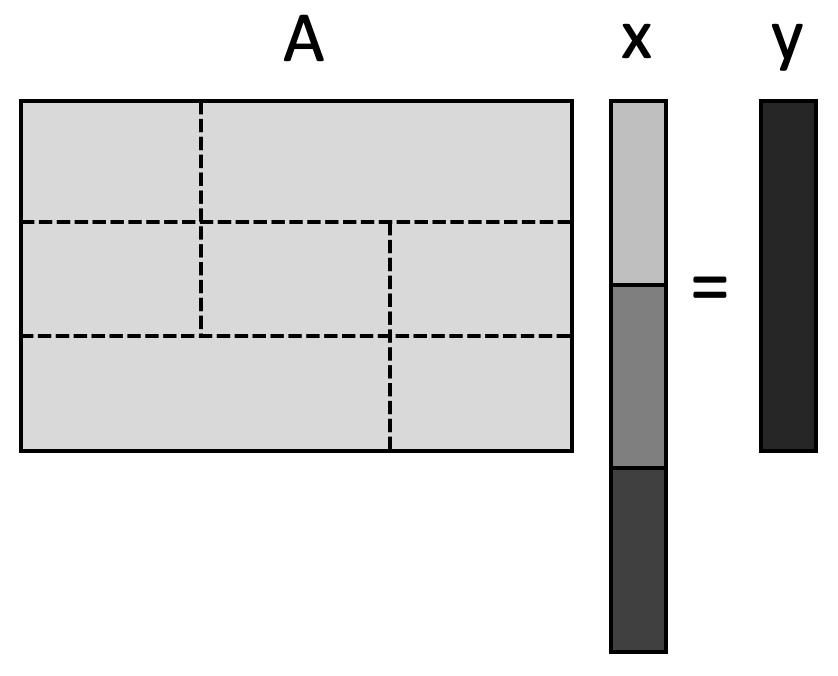}
         \caption{SpMV $y=Ax$.}
         \label{fig:spmv-whole}
     \end{subfigure}
     \hfill
     \begin{subfigure}[t]{0.56\textwidth}
         \centering
         \includegraphics[width=\textwidth]{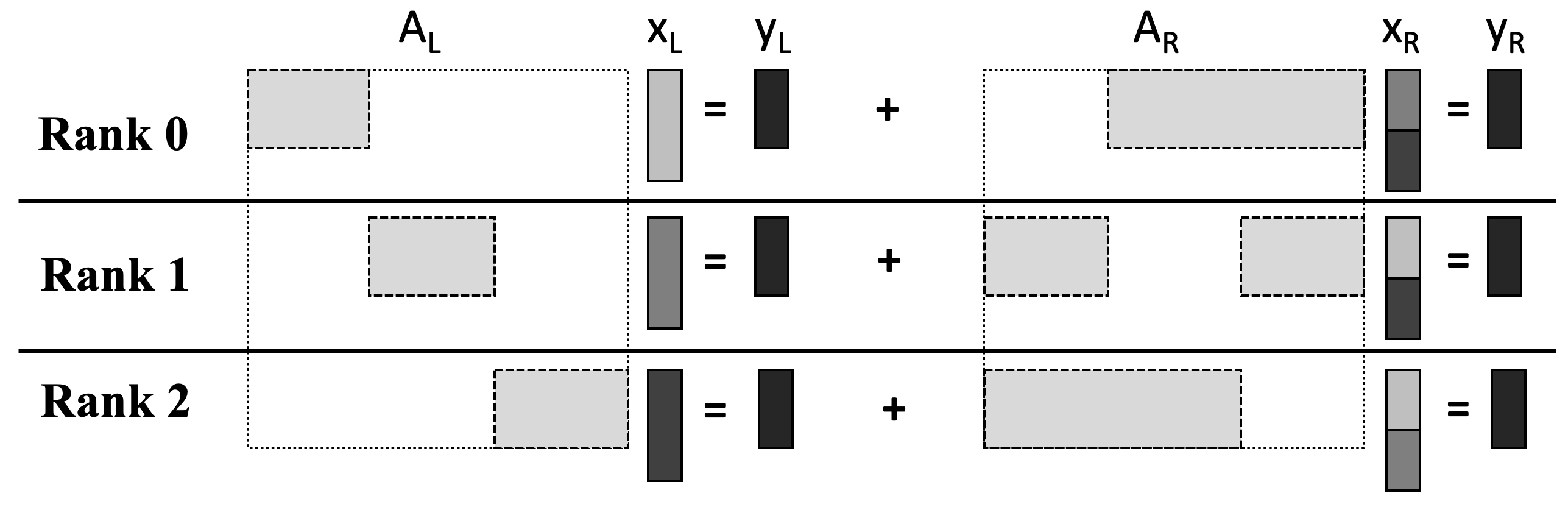}
         \caption{Diagram of an SpMV with row-wise partitions of $A$, $x$, and $y$ across three ranks.}
         \label{fig:spmv-mpi}
     \end{subfigure}
     \hfill
     \begin{subfigure}[t]{0.19\textwidth}
         \centering
         \includegraphics[width=\textwidth]{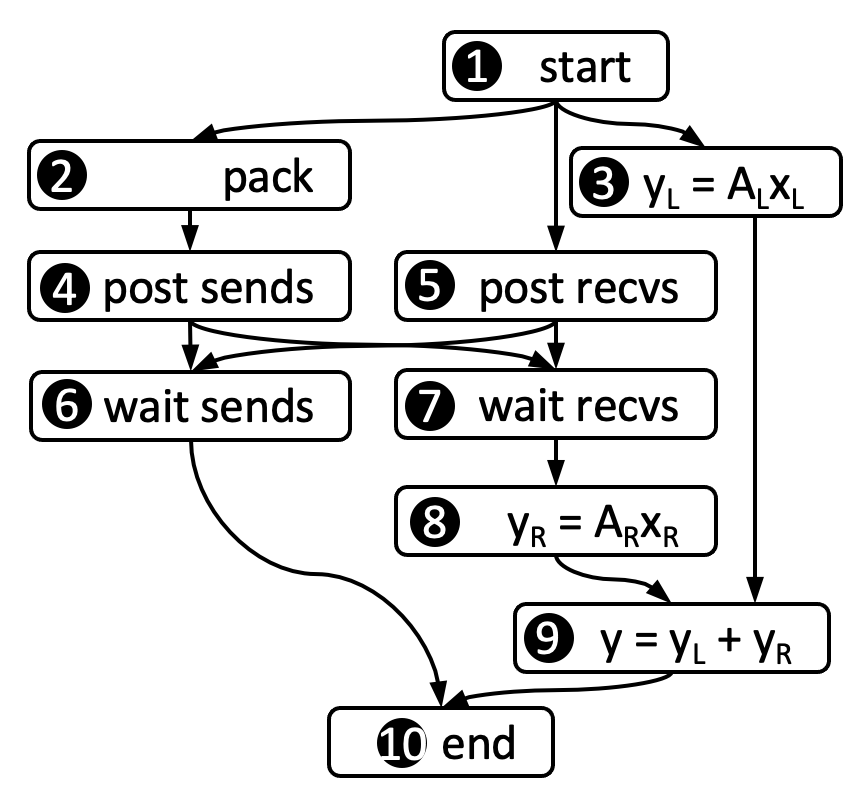}
         \caption{Directed acyclic graph of an SpMV program.}
         \label{fig:spmv-dag}
     \end{subfigure}
    \caption{Distributed sparse-matrix vector multiplication.}
        \label{fig:spmv}
\end{figure*}

The proposed system is demonstrated using a distributed sparse-matrix vector multiplication, summarized in Figure~\ref{fig:spmv}, on the Perlmutter platform, summarized in Table~\ref{tab:perlmutter}.
The SpMV is distributed across four ranks in a single node (Figure~\ref{fig:spmv-mpi}).
The matrix $A$ is a band-diagonal matrix with \num{150000} rows/columns, \num{1500000} non-zeros and a bandwidth of $\frac{150000}{4}$.
This bandwidth approximately balances the size of local and remote matrix multiplications (Sec.~\ref{sec:dag}).
The non-zeros are uniformly randomly distributed within the band.
GPU operations can be scheduled between two CUDA streams.

\begin{table}[ht]
\centering
\caption{
Perlmutter system description
}
\label{tab:perlmutter}
\begin{tabular}{c|c}
\textbf{Component} & \textbf{Description}  \\ \hline
CPU & AMD EPYC 7713 (64 cores, 2.0 GHz)\\
GPU & Nvidia A100 (PICe 4.0, 40GB) \\
OS & SUSE Linux Enterprise Server 15 SP2 \\
Kernel & 5.3.18 \\
GCC & 7.5.0 \\
MPI & Cray-MPICH 8.1.11.15 (MPICH 3.4a2) \\
CUDA & 11.4 \\
Nvidia Driver & 450.162 \\
Python & 3.9.4 \\
scikit-learn & 0.24.2 \\
scipy & 1.6.1 \\
\end{tabular}
\end{table}

\subsection {DAG Representation of CUDA + MPI operations}
\label{sec:dag}

Consider an arbitrary CUDA+MPI program $P$.
Program $P$ contains operations that can depend on each other.
In this sense, $P$ defines a graph $G_P$: the vertices of $G_P$ are the operations of $P$ and the edges of $G_P$ are the dependencies between them.
For example, if $P$ consists of two operations \ding{202} and \ding{203}. and the latter can start only after the former completes, then $G_P$ is \ding{202} $\rightarrow$ \ding{203}. 

To more precisely define ``operations," we include Table~\ref{tab:vertices}. An operation is either synchronous or asynchronous, and it can correspond to computation (e.g., CUDA kernel launches), communication (e.g., MPI point-to-point and collective functions), or other supporting instructions (e.g., $\texttt{MPI\_Wait}$, $\texttt{cudaDeviceSynchronize}$).
This is captures the predominant model for distributed GPU programming, where a CPU control thread offloads the bulk of the compute to asynchronous GPU operations, coordinated with asynchronous MPI communication, and interspersed with a small amount of synchronous CPU operations to marshal or organize data.

Our work pertains only to programs $P$ for which $G_P$ is a DAG (i.e., for which $G_P$ is directed and acyclic).
If $P$ has an unbounded loop, $G_P$ is not a DAG, but the system can be applied to the loop body in isolation, or a fixed number of unrolled iterations.

\begin{table}[ht]
\centering
\caption{
Types of DAG Vertices
}
\label{tab:vertices}
\resizebox{\linewidth}{!}{%
\begin{tabular}{c|l}
\textbf{Vertex Type} & \textbf{Description}   \\ \hline
CPU & A synchronous CPU operation \\
GPU & An asynchronous GPU operation not yet assigned to a stream \\
BoundGPU\textsubscript{s} & A GPU vertex assigned to execution stream \textit{s} \\
\end{tabular}
}
\end{table}

The dependence graph $G_P$ of our SpMV example is a DAG; see Figure~\ref{fig:spmv-dag}. 
Consider multiplication of sparse matrix $A$ with vector $x$ to produce vector $y$ shown in Figure~\ref{fig:spmv}.
A common distributed memory implementation evenly divides contiguous rows of $A$, $x$, and $y$ evenly across MPI ranks.
A rank's $y$ entries can then be computed as the sum of a ``local'' and ``remote'' matrix-vector multiplication $y_L = A_L x_L$ and $y_R = A_R x_R$, as in Figure~\ref{fig:spmv-mpi}.
$A_L$ has the column entries of $A$ that correspond to $x_L$, the locally-held rows of $x$; $A_R$ has the rest, corresponding to $x_R$, the rows of $x$ held by other ranks.  Thus, $A = A_L + A_R$.
$A_L x_L$ can be computed directly, but $A_R x_R$ must wait for $x_R$ to be assembled from the remote $x$ entries that correspond to non-zero columns in $A_R$.

In this formulation, $A$ is considered to be static, so the entries that make up $x_R$ are fixed.
The values in those entries are not considered to be static, so each rank must copy a subset of its $x_L$ entries into one buffer for each other rank (the \textit{Pack} vertex \ding{203}).
Communication is done with point-to-point \texttt{MPI\_Isend}s (\textit{PostSends} vertex \ding{205}) and \texttt{MPI\_Irecv}s (\textit{PostRecvs} vertex \ding{206}).

The granularity of the operations is flexible.
For example, SpMV could have been implemented with a set of parallel independent vertices for each separate pack and MPI\_Isend instead of collecting them into single \textit{Pack} and \textit{PostSends} vertices.
This finer granularity would eliminate false dependencies between packing for one rank and sending to another, e.g. not being able to send to rank 1 before the pack for rank 2 is completed.
The downside of this fine granularity is a larger space of implementations to search.

Artificial \textit{start} and \textit{end} vertices are added, where there must be a path from \textit{start} to each vertex and a path from each vertex to \textit{end}.
\textit{Start} serves as a single entry point for the program, and \textit{end} as an artificial CPU operation to ensure all program operations have completed before the program completes. 

A non-trivial program will have some degree of flexibility with respect to the actual order of operations.
For example, in Figure~\ref{fig:spmv-dag},
one valid execution sequence is  \ding{202}-\ding{203}-\ding{204}-\ding{205}-\ding{206}-\ding{207}-\ding{208}-\ding{209}-\ding{210}-\ding{211}.
Another is \ding{202}-\ding{206}-\ding{204}-\ding{203}-\ding{205}-\ding{208}-\ding{207}-\ding{209}-\ding{210}-\ding{211}.

\subsection {
CUDA+MPI Programs as a Sequential Decision Problem}
\label{sec:sequential-decision}

In the general case where some operations are asynchronous and may execute in parallel, it is not possible to evaluate the performance of a partial program sequence, nor is it possible to look at a set of operations with finished predecessors and choose the successor from among them that will ultimately result in the fastest execution.
Only with the entire program sequence in place can the performance be determined.

Recall that $G_P$ is a DAG whose vertices are the operations of a CUDA+MPI program $P$ and whose edges are the dependencies of these operations.
A topological traversal of $G_P$ specifies $P$, where all dependencies of a vertex are completed before the vertex is executed.
We address the following problem: Given $G_P$ and an objective function $f$, we seek a traversal of $G_P$ for which $f$ is minimal.
In other words, we seek the complete program which respects the operation dependencies in the DAG and minimizes some objective function (e.g. execution time, our specific function is described in Sec.~\ref{sec:selection}).
More formally, we seek a traversal $P^*$ that solves
\begin{align}\label{eq:opt}
  \text{minimize } f(P) \quad \text{subject to} \quad P \in \mathcal{P},
\end{align}
where $\mathcal{P}$ is the set of traversals of $G_P$. In the  SpMV example, $f(P)$ is the expected runtime of $P$, but a different $f$ could be of interest depending on the application.

Solving (\ref{eq:opt}) exactly is intractable for $G_P$ and $f$ that arise in practice.
We therefore reformulate (\ref{eq:opt}) as a sequential decision problem \cite{bertsekas2019reinforcement}.
The reformulation is equivalent to (\ref{eq:opt}) and is also intractable, but in the sequential decision paradigm, $P^*$ is constructed one vertex at a time, which is a natural way to construct a graph traversal.

To introduce the sequential decision alternative to (\ref{eq:opt}), we first describe its states.
Let $P_k$ be a \textit{prefix} of length $k$; $P_k$ is a list of $k$ vertices from $G_P$ that could later be extended into a complete traversal (i.e., to an element in $\mathcal{P}$).
Since the DAG $G_P$ has a finite $N$ vertices, the prefix $P_k$ should have at least $k\geq 0$ and no more than $N$ vertices (at which point it is a complete traversal from $\mathcal{P}$).

\subsection{Monte-Carlo Tree Search for CUDA+MPI Programs}
\label{sec:mcts}

To compute $P^*_k$ (the length $k$ prefix of an optimal program $P^*$), we must know the optimal performance of all programs beginning with all prefixes having length not greater than $k$. Deriving this information is infeasible in general. Were we to have it, however, computing $P^*_k$ from $P^*_{k – 1}$ would be trivial: we would append to $P^*_{k – 1}$ a vertex that preserves the best possible performance of programs beginning with $P^*_{k – 1}$ (which is $f(P^*)$ by definition). We refer the reader to \cite{bertsekas2019reinforcement,sutton2018reinforcement} for introductions to this sequential decision making paradigm.

Monte-Carlo tree search is a technique that approximates this exact yet infeasible construction of $P^*$ by estimating the optimal performance of the most impactful prefixes. It stochastically approximates $P^*$ while estimating the optimality of various $P_k$ along the way.
The state of the search is represented as a tree, where each tree node is an operation from the DAG and the ancestors of that node represent the prefix $P_k$ taken to get there.
The tree is built through four iterated phases: \textit{selection} chooses the subtree to explore, \textit{expansion} increases the information stored about that subtree, \textit{rollout} estimates optimality of $P_k$, and \textit{backpropagation} stores that information in the tree.
For MPI programs, MCTS is executed on only a single rank, with all ranks participating in the empirical measurement of the program.

\subsubsection{Selection}
\label{sec:selection}
A heuristic is used to favor exploring subtrees that feature larger observed performance ranges.
Starting with the root, children are recursively selected by maximizing an explore/exploit value.
The exploration value is defined as $ c \sqrt{\frac{\ln{N}}{n}}$,  where $c = \sqrt{2}$ is the exploration parameter, $N$ is the number of rollouts in which the tree node participated, and $n$ is the number of rollouts the child has participated in.
If the child subtree has been fully explored (all possible descendants have been benchmarked), the exploration value is set to negative infinity.
The exploitation value is given by
\[
V =
\begin{cases}
\frac{t^c_{max} - t^c_{min}}{t^p_{max} - t^p_{min}} & n \geq 2 \land N \geq 2 \\
1 & \text{otherwise}
\end{cases}
\]
where $t^c_{max}$ and $t^c_{min}$ are the maximum and minimum time observed in the child subtree, and similarly $t^p_{max}$ and $t^p_{min}$ for the parent subtree.
The intuition is to favor child nodes with times that represent greater coverage of the parent's execution times.
Since all child subtree times are contained in the parent subtree, $0 \leq V \leq 1$.
If a child or parent has not recorded two executions in its subtree, there is no basis for comparison and $1$ is used.

The combined value is the sum of the exploration and exploitation components and the largest value is selected.
The recursive search terminates at any node that has a child with no rollouts.

\subsubsection{Expansion}
A zero-rollout child of the node chosen by the selection phase is created.
This allows the search algorithm to retain additional information about this high-value region of the implementation space discovered during selection.
The selected node corresponds to a $P_k$, which is used to select the possible next operations based on the DAG.
The set of next operations is all vertices $v$ in $G_P$ not in $P_k$ and where all predecessors are in $P_k$, with two additional considerations.

First, any GPU-type vertices in $G_P$ must be converted to BoundGPU vertices by assignment to a CUDA stream.
In principle, each GPU vertex is replaced with one BoundGPU vertex for each stream; however, multiple available streams can introduce redundancy to $\mathcal{P}$.
For instance, if $P$ is defined on two equivalent streams $s_1$ and $s_2$, then the $P'$ that arises from swapping all assignments to $s_1$ and $s_2$ should attain the same value of $f$. To circumvent this redundancy, any children that represent equivalent $P_k$ under a stream bijection are pruned from the tree.

Second, the DAG does not require synchronization operations to be explicit.
Therefore, some prefixes may require synchronization operations before proceeding to the next vertex of $G_P$.
For example, in Figure~\ref{fig:spmv-dag}, edge \ding{203} $\rightarrow$ \ding{205}  requires a synchronization operation because the GPU kernel in \ding{203} must complete before the MPI operations in \ding{205} can start.
Table~\ref{tab:sync} summarizes when and how synchronization operations arise.
These synchronization operations depend on $P_k$, not the DAG, so they cannot be inserted in a preprocessing step.

\begin{table*}[ht]
\centering
\caption{
Inserting synchronization operations between vertices \textit{u}$\rightarrow$\textit{v} from the program DAG.
}
\label{tab:sync}
\resizebox{\textwidth}{!}{%
\begin{tabular}{c|c|c|p{3.5in}}
\textbf{\textit{u} Type} & \textbf{Inserted} & \textbf{\textit{v} Type} & \centering \textbf{Comment} \tabularnewline \hline
CPU & none & CPU or BoundGPU\textsubscript{i} & CPU vertices are synchronous, so \textit{u} is complete when \textit{v} starts \\
BoundGPU\textsubscript{i} & cudaEventRecord $\rightarrow$ cudaEventSync & CPU & BoundGPU operations are async, so must explicitly sync before CPU operation \\
BoundGPU\textsubscript{i} & none & BoundGPU\textsubscript{i} & Operations in the same stream are implicitly synchronized \\
BoundGPU\textsubscript{i} & cudaEventRecord $\rightarrow$ cudaStreamWaitEvent & BoundGPU\textsubscript{j} & Operations in different streams must be explicitly synchronized \\
\end{tabular}
}
\end{table*}

\subsubsection{Rollout}
\label{sec:mcts-rollout}
The performance of the subtree rooted at the node created in the expansion phase is estimated.
As described in Section~\ref{sec:sequential-decision}, the optimality of $P_k$ is not directly available.
Therefore, it is estimated by constructing a complete $P$, with prefix corresponding to the expanded node.
Recursively, random children are selected until the operation sequence is complete.

The generated $P$ is broadcast to all ranks.
Benchmarking is done in terms of \textit{measurements} and \textit{samples}.
Each sample is a single invocation of $P$, and each measurement corresponds to a single estimate of the time taken to execute $P$.
A measurement is generated by starting a timer and repeating $n_{samples}$ samples until a time $t_{measure}$ of \SI{0.01}{\second} has elapsed.
The time for $P$ is estimated as the maximum $\frac{t_{measure}}{n_{samples}}$ across all ranks.

The empirical performance measurements are recorded and stored alongside the selected sequence to use for rule generation (Section~\ref{sec:design-rule-generation}).
The nodes corresponding to this random rollout are constructed and added to the tree as well to retain their performance information.

\subsubsection{Backpropagation}
The results from the rollout are propagated to each ancestor node along the rollout path.
Each node tracks the fastest ($t_{min}$) and slowest ($t_{max}$) times observed during rollouts in the subtree rooted at that node.
If a rollout in the subtree produces an empirically measured time $t$, then each node along the path is updated as $t_{min} \leftarrow min(t, t_{min})$ and $t_{max} \leftarrow max(t, t_{max})$.

\section{Design Rule Generation}
\label{sec:design-rule-generation}

After MCTS has explored interesting regions of the implementation space, the empirical performance of those implementations is used to generate design rules for different performance classes.
Recall that during the MCTS rollout, the empirical time for each generated $P$ was recorded along with $P$ itself.
This records the performance of various points in the design space, biased towards regions of the design space where there is greater performance variability, i.e., where design decisions have a larger impact.

\subsection{Generating Class Labels}
\label{sec:class-labels}

\begin{figure}[ht]
\centering
\includegraphics[width=\linewidth]{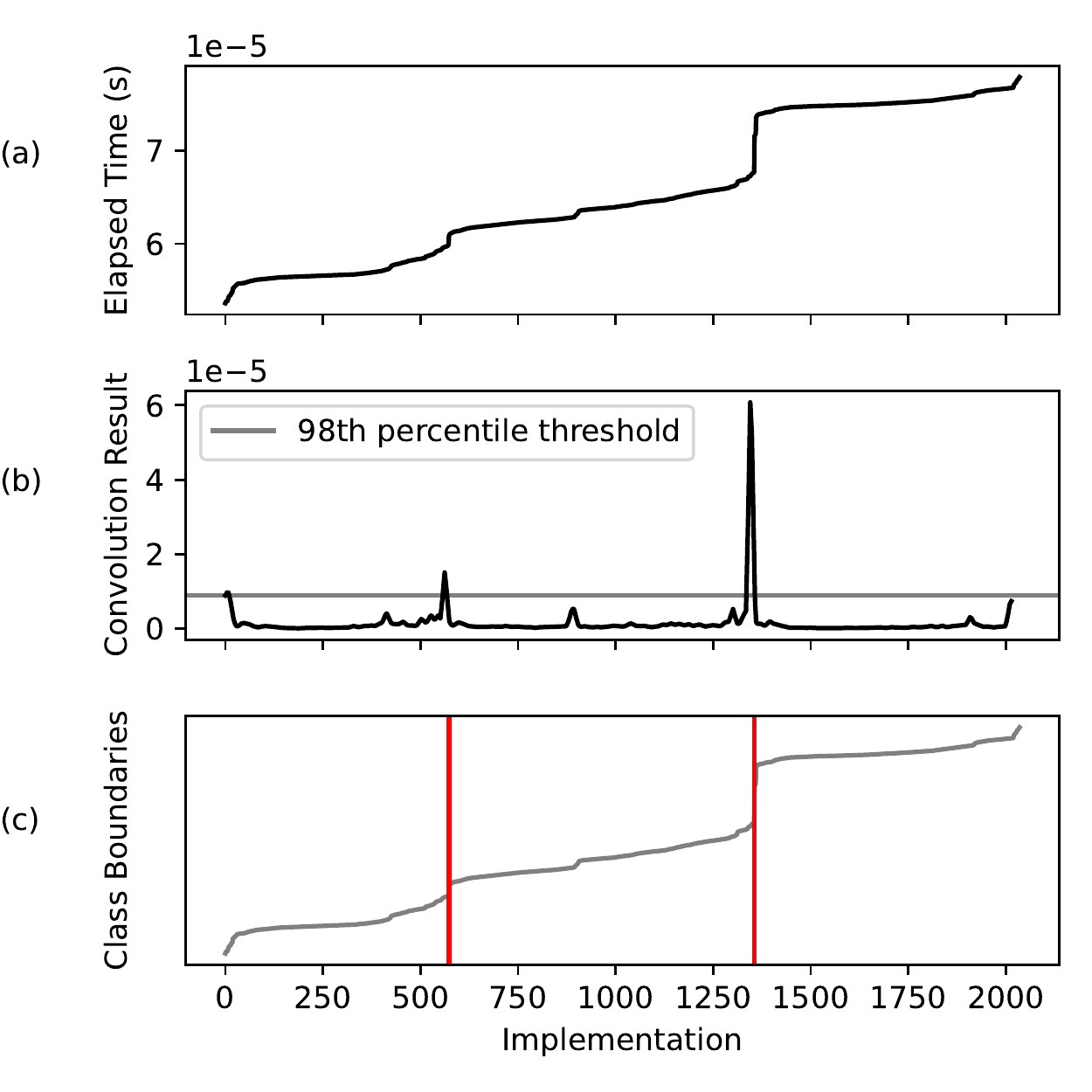}
\caption{
(a) the sorted empirical measurement data (fastest to slowest);
(b) the result of the convolution and a prominence threshold percentile to discard small peaks;
(c) the detected class boundaries overlaid on the original measurement data.
}
\label{fig:classes}
\end{figure}

The traversals explored during MCTS are grouped into classes according to their performance.
Figure~\ref{fig:classes} summarizes the automatic class labeling process.
First, the benchmark data is sorted least to greatest (Figure~\ref{fig:classes}a).
Then, the benchmark data is convolved with a step function with a radius $r$, given by
\[
k_m = 
\begin{cases}
-1 & -r \leq m \leq 0 \\
1 & 0 < m < r
\end{cases}
\]
where the convolution 
\[
\sum_{m=-r+1}^{r} k_m \times a_{i + m}
\]
is computed for all $i$ where $r < i < \text{length}(a)-r$,
i.e., only where the convolution kernel completely overlaps with the array $a$. 
The radius of the step function is set to 0.5\% (minimum 1) of the number of measurements to screen away small fluctuations in the performance.
This convolution produces the result in Figure~\ref{fig:classes}b.

The convolution result has a peak where the sorted performance measurements take a large increase.
Peaks are detected through a simple comparison with neighboring values~\cite{scipy-signal-findpeaks, scipy}.
Small peaks and corresponding small time increases can be screened away by choosing only the most prominent peaks.
In this work, only peaks above the 98th percentile of prominence are kept.
Thus, the number of performance classes need not be known a priori.
Each peak location is used as a boundary between performance classes (Figure~\ref{fig:classes}c).
Each traversal sequence is then labeled with the corresponding class.

\subsection{Generating Feature Vectors}
\label{sec:feature-vectors}

Each traversal is transformed into a fixed-length vector of \textit{ordering} and \textit{stream assignment} features.
An ordering feature is defined for each pairwise combination of traversal operations $u$ and $v$.
This feature is $1$ if $u$ appears in the traversal before $v$, and $0$ otherwise.
Similarly, a stream assignment feature is defined for each pairwise combination of BoundGPU operations.
This feature is $1$ if $u$ and $v$ occur in the same stream, and $0$ otherwise.
Many of these feature entries will have the same value for all traversals, e.g. $u$ will precede $v$ in all traversals if $u \rightarrow v$ in the program DAG.
Such features are removed from the resulting vectors as they provide no discriminatory power for different sequences.

\subsection{Decision Tree Training}
\label{sec:tree-training}

The scikit-learn python package~\cite{scikit-learn} is used to train a decision tree using the class labels and feature vectors.
scikit-learn implements the CART algorithm~\cite{breiman2017classification}.
Training involves several hyperparameters summarized in Table~\ref{tab:tree-params}.
\begin{table*}[ht]
\centering
\caption{
scikit-learn decision tree training parameters
}
\label{tab:tree-params}
\begin{tabular}{c|c|c}
\textbf{Parameter Name} & \textbf{Value} & \textbf{Comment} \\ \hline
\texttt{criterion} & \texttt{gini} & or \texttt{entropy}. \texttt{gini} is a simpler and faster caculation, no difference for test cases \\ \hline
\texttt{max\_depth} & chosen by Alg.~\ref{alg:dt-params} & Maximum allowable tree depth \\ \hline
\texttt{max\_leaf\_nodes} & chosen by Alg.~\ref{alg:dt-params} & Maximum number of leaf nodes in the tree \\ \hline
\texttt{class\_weight} & \texttt{balanced} & Weight all classes equally regardless of how many inputs have each label \\
\end{tabular}
\end{table*}

Ultimately, the paths to leaf nodes of the decision tree are used to determine discriminatory features of each class of sequences.
Therefore, a maximally-accurate decision tree is desired, without concern for overfitting.
The number of leaf nodes of the tree is initially set to the number of classes, and iteratively increased until classification error no longer shrinks.
Algorithm~\ref{alg:dt-params} produces the decision tree classifier \textit{clf}.
``train()'' is a function that takes \texttt{max\_leaf\_nodes} as an argument, and trains a tree with the provided \texttt{max\_leaf\_nodes} and \texttt{max\_depth} $=$ \texttt{max\_leaf\_nodes}$-1$ and returns an \texttt{(error}, \texttt{classifier)} tuple.

\begin{figure}[ht]
\centering
\includegraphics[width=\linewidth]{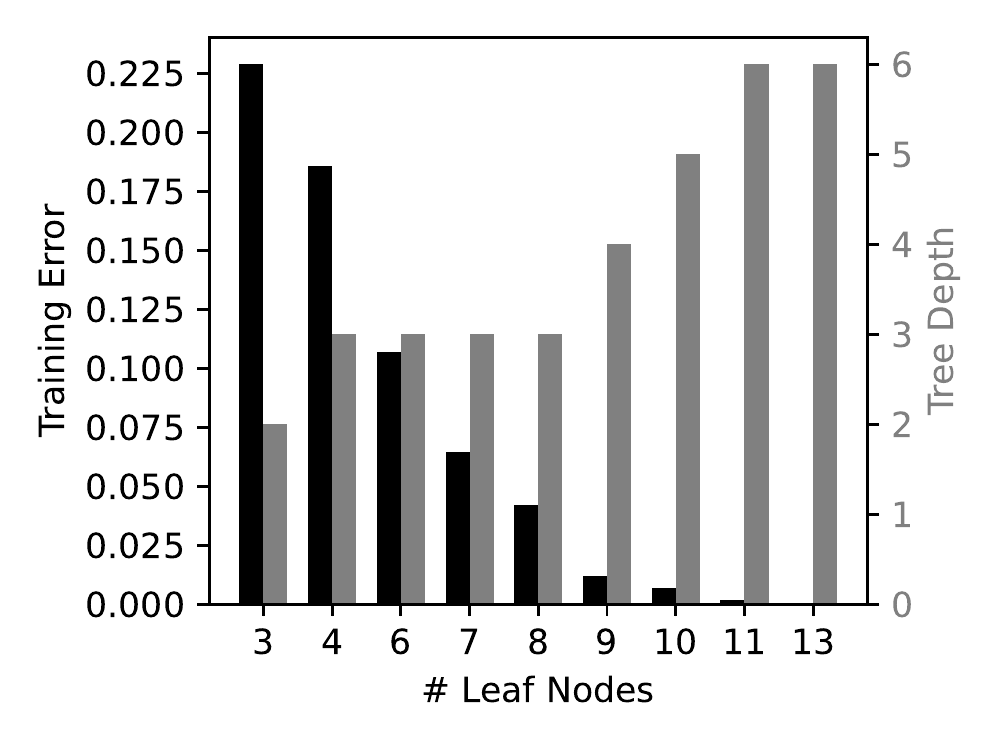}
\caption{
Training error and tree depth during decision tree hyperparameter search.
The smallest number of leaf nodes that minimizes the error is sought.
Maximum tree depth is restricted to one less than the number of leaf nodes, though the tree may not reach the allowable depth.
}
\label{fig:dt-hyperparams}
\end{figure}

\begin{algorithm}
\caption{Decision tree hyperparameter search producing classifier \textit{clf}.}
\label{alg:dt-params}
\begin{algorithmic}
\State $mln \gets 2$ \Comment{max\_leaf\_nodes}
\State $err \gets \infty$ \Comment{training error}
\State $cur, \mathit{clf} \gets \text{train}(mln)$ \Comment{error, classifier tuple}
\While{$cur < err$}
\State $err \gets cur$
\For{$i$ in 1 to 5} \Comment{try increasing tree size}
\State $\mathit{cur}, \mathit{nclf} \gets \text{train}(mln+i)$
\If{$cur < err$} \Comment{better hyperparameters}
    \State $\mathit{clf} \gets \mathit{nclf}$
    \State $\mathit{mln} \gets \mathit{mln} + i$
    \State break 
\EndIf
\EndFor
\EndWhile
\end{algorithmic}
\end{algorithm}
Figure~\ref{fig:dt-hyperparams} shows the result for the sample problem, ultimately settling on 13 nodes with a tree depth of 6.

\subsection{Rule Generation}
\label{sec:rule-generation}

\begin{figure*}[ht]
\centering
\includegraphics[width=\linewidth]{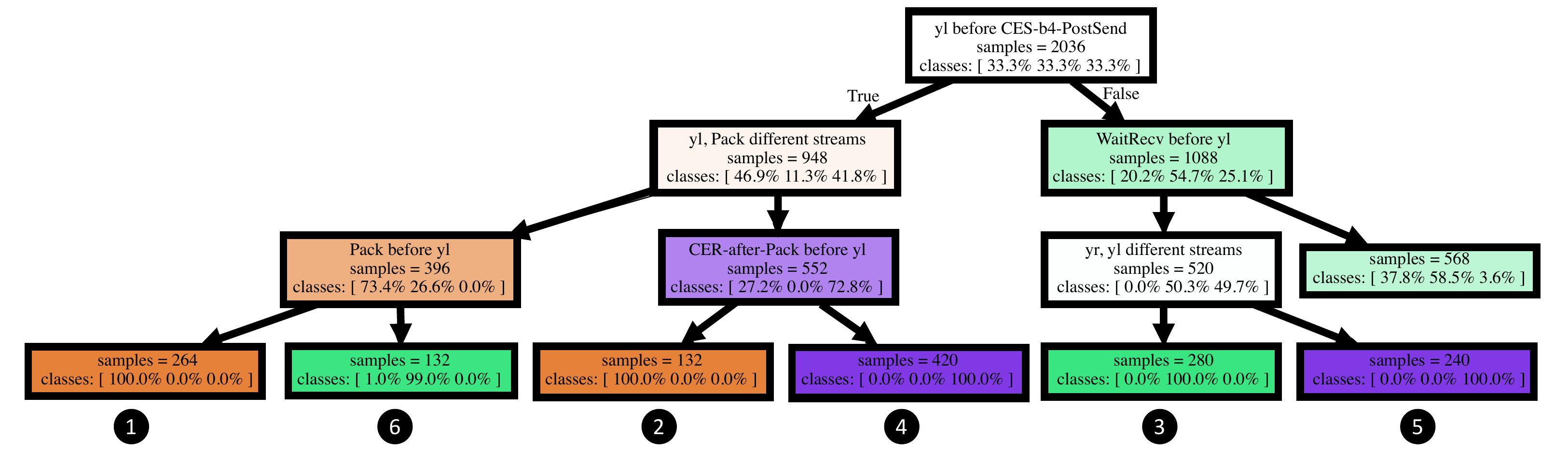}
\caption{
Decision tree with six leaf nodes and depth four for SpMV with \num{150000} non-zeros.
The paths to  \ding{202} and \ding{203} represent different sets of rules that place an implementation in the fastest performance class.
\ding{204} is the intermediate class, and \ding{205} and \ding{206} the slowest.
\ding{207} is a ruleset that contains samples from two classes, suggesting a deeper tree is needed.
The node colors signify which class assignment dominates the samples.
}
\label{fig:trained}
\end{figure*}

The design rules that define each performance class can be determined by all paths through the decision tree that arrive in a leaf node that contains that performance class.
Figure~\ref{fig:trained} shows an intermediate decision tree discovered during the iterative parameter search process in Section~\ref{sec:tree-training}. (The full tree is unwieldy to display, but the procedure described below is the same.)
``Samples'' refers to the number of training samples that fall into a particular box.
``Classes'' describes what proportion of the samples fall into each class.
The path from the root node to \ding{202} describes the design rules that will place an implementation into that leaf node.
For node \ding{202}, those rules are:
\begin{enumerate}
    \item ``y\textsubscript{L} before CES-b4-PostSend'': CES-b4-PostSend is an inserted (and automatically named) synchronization operation before PostSend. The local multiplication should occur before that inserted synchronization. 
    \item ``y\textsubscript{L}, Pack in different streams'': the local multiplication and pack kernels should be in different streams
    \item ``Pack before y\textsubscript{L}'': the Pack operation should be launched before the local multiplication
\end{enumerate}

Note that leaf node \ding{203} provides an alternate set of rules for ending up in the same performance class.
\ding{204} provides rules for implementations in the intermediate class, and \ding{205} and \ding{206} in the slowest class.
Finally, it is possible that the features cannot perfectly discriminate a class (\ding{207}).
In this case, it is because the intermediate tree was not allowed enough leaf nodes.

\section{Evaluation of MCTS}\label{sec:mcts-results}

Prior sections evaluated the techniques on an exhaustive search of the solution space for the provided SpMV.
The class labels and rules were generated from the complete 2036 possible traversals.
For larger applications, such a comprehensive search is not possible, and design rules must be generated from a subset of the search space.

\begin{table}[ht]
\centering
\caption{
Effect of MCTS iterations on labeling accuracy.
The given number of MCTS iterations is used to create class labels, each with an associated performance range.
That decision tree is used to classify all possible implementations
Accuracy is given as the proportion of implementations with performance that falls within the label's range, i.e., how well using only a subset generalized to the entire space.
}
\label{tab:exp2}
\begin{tabular}{c|c|c|c|c|c}
\textbf{MCTS Iterations} & 50 & 100 & 200 & 400 & 2036  \\ \hline
\textbf{Class Accuracy} & 0.75 & 0.83 & 0.96 & 0.99 & 1.0
\end{tabular}
\end{table}

The effect of MCTS are explored by examining how the discovered design rules change as more of the search space is explored.
Table~\ref{tab:exp2} summarizes the quality of the design rules depending on how many MCTS iterations are allowed.
The provided number of MCTS iterations executes, yielding a reduced set of traversals and corresponding execution times.
Features and class labels are generated from the subset of samples.
Each class defines a performance range, from the slowest sample in the class to the fastest.
The entire space of 2036 implementations is then classified using the rules derived from the reduced subset, and the proportion that fall within the performance range is reported.
As the number of MCTS iterations increases, the accuracy of design rules for classifying the design space increases, up to nearly 100\% at 200 iterations.

A qualitative way of examining the results is presented in Tables~\ref{tab:class1}, \ref{tab:class2}, and \ref{tab:class3}, which summarize the design rules generated for each performance class.
Each column of the table displays the generated rulesets that discriminate a given performance class, and each cell in the column corresponds to a different path through the decision tree to that class.
When there are multiple paths to the same performance class, the cells in a column may be mutually contradictory.
For example, the first and third cell in the \num{2036} column of Table~\ref{tab:class1} have different rules for whether the Pack and y\textsubscript{L} operations should be in different streams.
However, each ruleset prescribes the necessary conditions for an implementation to be in a performance class; as long as all rules in a given ruleset are followed, other decisions do not matter.
The cells are sorted top-to-bottom by the number of training samples that followed those rules.
For brevity of presentation, only the three rulesets with the most samples are shown for each case.

The rules generated for the full SpMV traversal (rightmost column, having 2036 implementations) are taken as the canonical, accurate rules since they reflect classification of all possible implementations.
The cells are annotated with information about their consistency with these canonical rules.
In an ideal system, each cell in a column would correspond directly to one cell in column \num{2036} (though not necessarily the entry in the same row); however, two kinds of inconsistencies are observed.

First, a ruleset may be overconstrained, or consistent with the canonical rules but with additional harmless restrictions.
This is highlighted in \blue{blue} in each cell.
Table~\ref{tab:class1} shows that for the fastest performance class, the rules discovered by fewer MCTS iterations tend to be overconstrained, with almost all rulesets featuring at least one spurious rule.
Second, a ruleset may be underconstrained; i.e., it does not restrict the order and assignment of operations sufficiently to fit within the comprehensive search rules.
The two slower performance classes (Tables~\ref{tab:class2} and \ref{tab:class3}) suffer from this problem.  Relatively few rules are completely accurate; most are missing at least one constraint.
This is annotated in \red{red} in each cell.
Although most cells have an error, they are typically ``almost'' correct when compared to the rules generated from the full traversal (column \num{2036}).

There are two ways to use these rules.
First, program implementors may take any ruleset that corresponds to the desired performance class and follow the rules in their implementation.
Doing so will ensure the performance of the implementation falls within that class.
Second, implementors can analyze the rules to gain deeper insight into the properties of the implementation.

Additional study is necessary to determine why class 1 behaves so distinctly from classes 2 and 3, and why the quantitative accuracy is so high (Figure~\ref{tab:exp2}) despite the inconsistencies in textual rules.
It may be that the inaccuracies in the rulesets are enough to prevent a perfect match with the rulesets for the exhaustive search, but are not so significant that they push a program $P$ into the wrong class very often.

\begin{table*}[ht]
\centering
\caption{
Design rules that define performance class 1 generated for various MCTS iterations.
The three rulesets generated for 2036 iterations are taken as canonical.
All rulesets generated with fewer search iterations are consistent with one of the canonical rulesets.
Rules highlighted in \blue{blue} are extraneous, but do not violate a ruleset. 
}
\label{tab:class1}
\resizebox{\textwidth}{!}{%
\begin{tabular}{c|c|c|c|c}
\textbf{50} & \textbf{100} & \textbf{200} & \textbf{400} & \textbf{2036} \\ \hline
\makecell[tl]{y\textsubscript{L} before CES-b4-PostSend \\ y\textsubscript{L} different stream than Pack \\ Pack before y\textsubscript{L}} 
& \makecell[tl]{y\textsubscript{L} before CES-b4-PostSend \\ y\textsubscript{L} different stream than Pack \\ Pack before y\textsubscript{L} \\ \blue{y\textsubscript{L} before WaitSend}} 
& \makecell[tl]{y\textsubscript{L} before CES-b4-PostSend \\ y\textsubscript{L} different stream than Pack \\ Pack before y\textsubscript{L} \\ \blue{y\textsubscript{L} before WaitSend}} 
& \makecell[tl]{y\textsubscript{L} before WaitRecv  \\ PostSend before y\textsubscript{L}  \\ Pack before y\textsubscript{L} \\ CER-after-Pack before y\textsubscript{L}  \\ y\textsubscript{L} before WaitSend  \\  PostRecv before CES-b4-PostSend} 
& \makecell[tl]{y\textsubscript{L} before CES-b4-PostSend \\ y\textsubscript{L} different stream than Pack \\ Pack before y\textsubscript{L}} 
\\ \hline
\makecell[tl]{y\textsubscript{L} before CES-b4-PostSend \\ Pack same stream as y\textsubscript{L} \\ CER-after-Pack before y\textsubscript{L}} 
& \makecell[tl]{y\textsubscript{L} before CES-b4-PostSend \\ Pack same stream as y\textsubscript{L} \\ CER-after-Pack before y\textsubscript{L} \\ \blue{y\textsubscript{L} before WaitSend} \\ \blue{y\textsubscript{L} before WaitRecv} \\ } 
& \makecell[tl]{CES-b4-PostSend before y\textsubscript{L} \\ y\textsubscript{L} before WaitRecv \\ PostSend before y\textsubscript{L} \\ y\textsubscript{L} before WaitSend \\ \blue{Pack before y\textsubscript{L}} \\ \blue{PostRecv before CES-b4-PostSend}} 
& \makecell[tl]{y\textsubscript{L} before WaitRecv \\ Pack before y\textsubscript{L} \\ CER-after-Pack before y\textsubscript{L} \\ y\textsubscript{L} before WaitSend \\ y\textsubscript{L} before PostSend \\ y\textsubscript{L} before CES-b4-PostSend} 
& \makecell[tl]{CES-b4-PostSend before y\textsubscript{L} \\ y\textsubscript{L} before WaitRecv \\ PostSend before y\textsubscript{L} \\ PostRecv before PostSend \\ y\textsubscript{L} before WaitSend} 
\\ \hline
\makecell[tl]{} 
& \makecell[tl]{y\textsubscript{L} before WaitSend \\ y\textsubscript{L} different stream than Pack \\ Pack before y\textsubscript{L} \\ CES-b4-PostSend before y\textsubscript{L} \\ \blue{PostSend before y\textsubscript{L}} \\ y\textsubscript{L} before WaitRecv \\ PostRecv before PostSend} 
& \makecell[tl]{y\textsubscript{L} before CES-b4-PostSend \\ Pack same stream as y\textsubscript{L} \\ CER-after-Pack before y\textsubscript{L} \\ \blue{y\textsubscript{L} before WaitSend} \\ \blue{Pack before y\textsubscript{L}}} 
& \makecell[tl]{\blue{y\textsubscript{L} before WaitRecv} \\ Pack before y\textsubscript{L} \\ y\textsubscript{L} before CER-after-Pack \\ y\textsubscript{L} different stream than Pack}
& \makecell[tl]{y\textsubscript{L} before CES-b4-PostSend \\ Pack same stream as y\textsubscript{L} \\ CER-after-Pack before y\textsubscript{L}}

\end{tabular}
}
\end{table*}

\begin{table*}[ht]
\centering
\caption{
Generated rules for class 2 (Same format as Table~\ref{tab:class1}).
}
\label{tab:class2}
\resizebox{\textwidth}{!}{%
\begin{tabular}{c|c|c|c|c}
\textbf{50} & \textbf{100} & \textbf{200} & \textbf{400} & \textbf{2036} \\ \hline
\makecell[tl]{\blue{y\textsubscript{L} different stream than Pack} \\ y\textsubscript{L} before Pack \\ \red{insufficient rules}} 
& \makecell[tl]{y\textsubscript{L} before WaitSend \\ Pack different stream than y\textsubscript{L} \\ \blue{y\textsubscript{L} before Pack} \\ \red{insufficient rules} } 
& \makecell[tl]{WaitSend before y\textsubscript{L} \\ yr different stream than y\textsubscript{L} \\ \red{insufficient rules}} 
& \makecell[tl]{WaitRecv before y\textsubscript{L} \\ yr different stream than y\textsubscript{L} \\ \red{insufficient rules}} 
& \makecell[tl]{CES-b4-PostSend before y\textsubscript{L} \\ WaitRecv before y\textsubscript{L} \\ yr different stream than y\textsubscript{L}} 
\\ \hline
\makecell[tl]{\blue{y\textsubscript{L} different stream than Pack}  \\ \blue{Pack before y\textsubscript{L}} \\ CES-b4-PostSend before y\textsubscript{L} \\ y\textsubscript{L} before yr \\ \red{insufficient rules}} 
& \makecell[tl]{y\textsubscript{L} before WaitSend \\ \blue{y\textsubscript{L} different stream than Pack} \\ Pack before y\textsubscript{L} \\ CES-b4-PostSend before y\textsubscript{L} \\ y\textsubscript{L} before PostSend \\ \red{insufficient rules}} 
& \makecell[tl]{y\textsubscript{L} before WaitSend \\ \blue{y\textsubscript{L} before Pack} \\ \blue{Pack different stream than y\textsubscript{L}} \\ \blue{PostRecv before Pack} \\ \red{insufficient rules}} 
& \makecell[tl]{y\textsubscript{L} before WaitSend \\ \blue{y\textsubscript{L} before Pack} \\ \blue{Pack different stream than y\textsubscript{L}} \\ \blue{PostRecv before Pack} \\ \red{insufficient rules}} 
& \makecell[tl]{CES-b4-PostSend before y\textsubscript{L} \\ y\textsubscript{L} before WaitRecv \\ y\textsubscript{L} before PostSend} 
\\ \hline
\makecell[tl]{\blue{Pack same stream as y\textsubscript{L}} \\ CER-after-Pack before y\textsubscript{L} \\ CES-b4-PostSend before y\textsubscript{L} \\ \red{insufficient rules}} 
& \makecell[tl]{y\textsubscript{L} before WaitSend \\ \blue{Pack same stream as y\textsubscript{L}} \\ \blue{CER-after-Pack before y\textsubscript{L}} \\ y\textsubscript{L} before WaitRecv \\ CES-b4-PostSend before y\textsubscript{L} \\ y\textsubscript{L} before PostSend} 
& \makecell[tl]{CES-b4-PostSend before y\textsubscript{L} \\  y\textsubscript{L} before PostSend \\ \blue{y\textsubscript{L} before WaitSend} \\ \blue{Pack before y\textsubscript{L}} \\ \red{insufficient rules}} 
& \makecell[tl]{CES-b4-PostSend before y\textsubscript{L} \\ y\textsubscript{L} before WaitRecv \\ y\textsubscript{L} before PostSend \\ \blue{Pack before y\textsubscript{L}} \\ \blue{CER-after-Pack before y\textsubscript{L}} \\ \blue{y\textsubscript{L} before WaitSend}} 
& \makecell[tl]{CES-b4-PostSend before y\textsubscript{L} \\ y\textsubscript{L} before WaitRecv \\ PostSend before y\textsubscript{L} \\ PostSend before PostRecv \\ y\textsubscript{L} before WaitSend} 
\end{tabular}
}
\end{table*}

\begin{table*}[ht]
\centering
\caption{
Generated rules for class 3 (Same format as Table~\ref{tab:class1}).
}
\label{tab:class3}
\resizebox{\textwidth}{!}{%
\begin{tabular}{c|c|c|c|c}
\textbf{50} & \textbf{100} & \textbf{200} & \textbf{400} & \textbf{2036} \\ \hline
\makecell[tl]{Pack same stream as y\textsubscript{L} \\ y\textsubscript{L} before CER-after-Pack \\ \red{insufficient rules}} 
& \makecell[tl]{\blue{y\textsubscript{L} before WaitSend} \\ Pack same stream as y\textsubscript{L} \\ y\textsubscript{L} before CER-after-Pack} 
& \makecell[tl]{\blue{y\textsubscript{L} before WaitSend} \\ \blue{y\textsubscript{L} before Pack} \\ Pack same stream as y\textsubscript{L}} 
& \makecell[tl]{WaitRecv before y\textsubscript{L} \\ yr same stream as y\textsubscript{L}} 
& \makecell[tl]{y\textsubscript{L} before CES-b4-PostSend \\ Pack same stream as y\textsubscript{L} \\ y\textsubscript{L} before CER-after-Pack} 
\\ \hline
\makecell[tl]{CES-b4-PostSend before y\textsubscript{L}  \\  WaitRecv before y\textsubscript{L} \\ \blue{Pack before y\textsubscript{L}}  \\ \blue{Pack same stream as y\textsubscript{L}} \\ \red{insufficient rules}} 
& \makecell[tl]{WaitSend before y\textsubscript{L} \\ yr same stream as y\textsubscript{L} \\ \red{insufficient rules}} 
& \makecell[tl]{WaitSend before y\textsubscript{L} \\ yr same stream as y\textsubscript{L} \\ \red{insufficient rules}} 
& \makecell[tl]{\blue{y\textsubscript{L} before WaitRecv} \\ \blue{y\textsubscript{L} before Pack} \\ Pack same stream as y\textsubscript{L}} 
& \makecell[tl]{CES-b4-PostSend before y\textsubscript{L} \\ WaitRecv before y\textsubscript{L} \\ yr same stream as y\textsubscript{L}} 
\\ \hline
\makecell[tl]{} 
& \makecell[tl]{CER-after-Pack before y\textsubscript{L} \\ WaitRecv before y\textsubscript{L} \\ \blue{y\textsubscript{L} before WaitSend} \\ \blue{Pack same stream as y\textsubscript{L}} \\ \red{insufficient rules}} 
& \makecell[tl]{y\textsubscript{L} before CES-b4-PostSend \\ Pack same stream as y\textsubscript{L} \\ y\textsubscript{L} before CER-after-Pack \\ \blue{y\textsubscript{L} before WaitSend} \\ \blue{Pack before y\textsubscript{L}}} 
& \makecell[tl]{y\textsubscript{L} before WaitRecv  \\ y\textsubscript{L} before CER-after-Pack \\ Pack same stream as y\textsubscript{L} \\ \blue{Pack before y\textsubscript{L}}  \\ \red{insufficient rules}} 
& \makecell[tl]{CES-b4-PostSend before y\textsubscript{L} \\ y\textsubscript{L} before WaitRecv \\ PostSend before y\textsubscript{L} \\ PostRecv before PostSend \\ WaitSend before y\textsubscript{L} \\ yr same stream as y\textsubscript{L}} 
\\ \hline
\makecell[tl]{} 
& \makecell[tl]{} 
& \makecell[tl]{CES-b4-PostSend before y\textsubscript{L} \\ WaitRecv before y\textsubscript{L} \\ yr same stream as y\textsubscript{L} \\ \blue{y\textsubscript{L} before WaitSend} \\ \blue{Pack before y\textsubscript{L}} \\  \blue{PostSend before y\textsubscript{L}}  } 
& \makecell[tl]{y\textsubscript{L} before WaitRecv  \\ CER-after-Pack before y\textsubscript{L} \\ WaitSend before y\textsubscript{L} \\ yr same stream as y\textsubscript{L} \\ \blue{Pack before  y\textsubscript{L}} \\ \red{insufficient rules}} 
& \makecell[tl]{CES-b4-PostSend before y\textsubscript{L} \\ y\textsubscript{L} before WaitRecv \\ PostSend before y\textsubscript{L} \\ PostSend before PostRecv \\ WaitSend before y\textsubscript{L} \\ yr same stream as y\textsubscript{L}} 
\end{tabular}
}
\end{table*}

\section{Future Work}

Substantial work is required to increase the capabilities of the system, as well as understand the quality of the results.
Results have been presented for only a single program (SpMV) on a single platform (Perlmutter with two streams per GPU) with a single input (the synthetic banded matrix).
Initial results suggest that the prototype is working well, but examination across additional platforms, programs, and inputs is required.
A natural extension is to generate rules that generalize across inputs.
This extension requires changes to the feature-vector generation to include features that discriminate between inputs.
The work is currently being extended to 3D halo-exchange communication modeling fine-grained communication operations in each dimension.
Also of interest is extending resource assignment to include multiple GPUs or NUMA nodes, instead of solely GPU streams.

It is clear that the decision tree hyperparameters affect the generated rules and their interpretability.
Tables~\ref{tab:class1}, \ref{tab:class2} and \ref{tab:class3} already suggest that interpreting the design rules may be challenging.
Currently, the most discriminating decision tree is sought; however, that objective may negatively impact the complexity and number of generated rules.
Additional work in feature vector generation and decision tree training parameters may increase the quality of the generated rules.

Finally, the MCTS search strategy preferentially explored design-space areas with high performance impact.
Other MCTS strategies should be considered, at least as a baseline for comparison.
For example, a search strategy that randomly samples the design space could be used to show that the current strategy indeed produces better results.

\section{Conclusion}

This work presents the design and initial evaluation of a system to generate design rules for CUDA + MPI operations.
It takes as input a DAG representation of the program, which is used in a Monte-Carlo tree search exploring performance-sensitive regions of the design space.
Performance classes and implementation features are derived from the results of the design space search, and then converted to textual rules.
Program developers can use these rules to  understand which aspects of their program most impact the program's performance.

Evaluation is limited to a specific CUDA+MPI sparse-matrix vector multiplication; additional work is needed to verify and extend the functionality of the system.
Of particular note is modifying the system to support generalized design rules across multiple inputs, and improving the rule generation and interpretibility.
For the limited scope of the current evaluation, the prototype produces promising results.

\section*{Acknowledgment}

Sandia National Laboratories is a multimission laboratory managed and operated by National Technology \& Engineering Solutions of Sandia, LLC, a wholly owned subsidiary of Honeywell International Inc., for the U.S. Department of Energy’s National Nuclear Security Administration under contract DE-NA0003525.
This paper describes objective technical results and analysis. Any subjective views or opinions that might be expressed in the paper do not necessarily represent the views of the U.S. Department of Energy or the United States Government.  

This work is supported by the U.S. Department of Energy, Office of Science, Office of Advanced Scientific Computing Research, Scientific Discovery through Advanced Computing (SciDAC) program through the FASTMath Institute.
This research used resources of the National Energy Research Scientific Computing Center (NERSC), a U.S. Department of Energy Office of Science User Facility located at Lawrence Berkeley National Laboratory, operated under Contract No. DE-AC02-05CH11231 using NERSC award ERCAP0019623.

The authors thank I-Hsin Chung of IBM T. J. Watson Research for his time and guidance.

\bibliographystyle{IEEEtran}
\bibliography{IEEEabrv, main} 

\end{document}